\documentclass[twocolumn,showpacs,preprintnumbers,amsmath,amssymb,prb]{revtex4}

\usepackage{graphicx}
\usepackage{dcolumn}
\usepackage{bm}

\begin{document}

\preprint{  }

\title{Evidence for 
Direct and Indirect Gap in FeSi\\ 
from Electron Tunneling Spectroscopy}

\author{M. Bat$\!$'kov\'a }   
\email{batkova@saske.sk}  
 \affiliation{Institute  of  Experimental  Physics,
 Slovak   Academy  of Sciences, Watsonova 47,
 040~01~Ko\v {s}ice, Slovakia}           
\author{I. Bat$\!$'ko}%
 \affiliation{Institute  of  Experimental  Physics,
 Slovak   Academy  of Sciences, Watsonova 47,
 040~01~Ko\v {s}ice, Slovakia}
\author{M. Mihalik}
\affiliation{Institute  of  Experimental  Physics,
 Slovak   Academy  of Sciences, Watsonova 47,
 040~01~Ko\v {s}ice, Slovakia}

\date{\today}

\begin{abstract}
 We report electron tunneling spectroscopy studies 
			on  single crystalline FeSi sample performed for the case 
			of homogeneous tunnel junction (TJ) contacts and for the case of counter electrode made from Pt-Rh alloy. 
Our results reveal that while the tunneling spectroscopy in the configuration with Pt-Rh tip is preferably sensitive to the
		 \mbox{\emph{d-}partial} density of states (DOS) and to the indirect energy gap, 
		 the \mbox{FeSi-FeSi} type of TJ yields the spectroscopic information on the	
		 	\mbox{\emph{c-}partial} DOS and on the direct gap in FeSi. \end{abstract}

\pacs{71.27.+a, 71.28.+d, 75.30.Mb, 73.40.Gk }
\maketitle

\hyphenation{Czoch-ral-ski}

\section{INTRODUCTION}
Cubic compound FeSi is well known by the unusual physical properties
		  arising from its unconventional band structure.
Despite the fact that FeSi is a \emph{d-}transition metal compound, 
			it shares certain characteristic 
			features with rare earth hybridization gap 
			semiconductors such as SmB$_6$, YbB$_{12}$ or Ce$_3$Bi$_4$Pt$_3$, 
			in which the hybridization of \emph{f-} and conduction electrons
			is believed to take place.
As a rule, FeSi is placed into the same special group 
				of heavy electron systems 
				with a narrow gap, named 
``Kondo insulators'' or ``Kondo semiconductors''
				 \cite{Degiorgi,Riseborough}.
 
According to transport and optical studies,
		 at high temperatures FeSi can be characterized as a dirty metal \cite{Hunt, 								Schlesinger}. 
As the temperature falls to around 300~K, it shows a metal 
			to semiconductor crossover \cite{Sales} and  behaves
			as a narrow gap semiconductor below $\sim$100~K 
			\cite{Sales,Mihalik,Paschen}.
A value of the activation energy $\Delta$     			
		  differs from about 50~meV to $\sim$110~meV upon the used
		  experimental method \cite{Schlesinger,Sales,Mihalik,Paschen,Fath,Aarts,Samuely}.
At the lowest temperatures FeSi has a metallic character 
			\cite{Hunt,Paschen,Degiorgi95}.  				  	
	  		   	
Several different theoretical approaches have been suggested 
			to explain striking properties and an origin of the energy gap in FeSi \cite{Mattheiss,Mandrus,Jarlborg,Rozenberg}.		 
A simple physical picture 
			assumes that instead of \emph{f-}electrons in the rare earth semiconductors, 
			a set of rather localized \emph{d-}orbitals having a strong on-site repulsion, 
			hybridizes in FeSi with a	broad itinerant conduction band of noninteracting 
			\emph{c-}electrons forming in this 
 			way two bands separated by the gap \cite{Degiorgi,Sales,Rozenberg}.
According to the theory of Rozenberg, Kotliar and Kajueter \cite{Rozenberg}, 
				the \emph{d-}electron partial DOS shows opening 
				of an indirect gap, $\Delta_{ind}$,
				accompanied by the formation of strong and narrow quasiparticle bands 
				visible as two symmetric peaks on either side of the Fermi level.
A direct gap, $\Delta_{dir}$, in \emph{c-}electron partial DOS is 
				merely a dip without peaks in the framework of this model \cite{Rozenberg}.
Moreover it is expected that $\Delta_{ind}<< \Delta_{dir}$ \cite{Rozenberg}. 
In accordance with  number of published experimental results, it appears that 
the magnetic gap probed by magnetic susceptibility \cite{Schlesinger,Sales,Mihalik,Paschen} and 		optical reflectivity  \cite{Mihalik,Paschen,Samuely} measurements is the larger direct gap,
and that the transport gap 	
			reflected in resistivity \cite{Schlesinger,Sales,Mihalik,Paschen,Samuely}, point contact 	spectroscopy \cite{Mihalik,Samuely}
  		and tunneling spectroscopy \cite{Fath} measurements is the smaller indirect gap. 
In this paper we show for the first time that tunneling spectroscopy is sensitive to 	
	     \emph{both} electron subsystems, and so that it can probe not only the indirect           but also the direct gap in FeSi.   

\section{EXPERIMENT}				  	
The samples used for experiments were pieces 
			of the same FeSi single crystal grown from the melt by the	``tri-arc''
			Czochralski technique, 
			for which magnetic, transport, 
			photo-emission, point-contact spectroscopy,
 			as well as infrared studies were reported previously \cite{Mihalik,Samuely,Breuer}.
The ratio of low temperature to room temperature resistivity,  $\rho(5~K)/\rho(300~K)=5\times10^3$ 
    is an indication for high quality sample \cite{Samuely}.
In addition, the electron probe microanalysis did not reveal 
    	a presence of any second phase \cite{Mihalik}.
As reported before, the value of the transport gap obtained from 
			resistivity data is $\Delta_{t}\sim$56~meV, 
			while the magnetic susceptibility studies indicate 
			the gap of $\Delta_{m}\sim$95~meV or $\sim$103~meV, 
			depending on the  fitting method used \cite{Mihalik}.		
		
The tunneling measurements 
		were performed 	by the scanning tunneling spectroscopy approach 
	using mechanically controlled 
	TJs in	two different configurations: 
(i) with Pt-Rh tip as a counter electrode and, for the first time as we know, 
	(ii) with the tip made from a piece of the same FeSi single crystal
	 as the studied sample.	
The Pt-Rh tip was prepared from commercially
			available Pt$_{0.9}$Rh$_{0.1}$ wire.
Differential conductance, $dI/dV$, was numerically calculated from the measured 
		current-voltage 
		characteristics.
It should be mentioned that due to the formation of an oxide layer on the FeSi surface, 
			the TJ electrodes were in mechanical contact, 
			separated just by this native insulating surface layer.
		
 \begin{figure}[!tb]
\begin{center}
\resizebox{0.99\columnwidth}{!}{%
  				\includegraphics{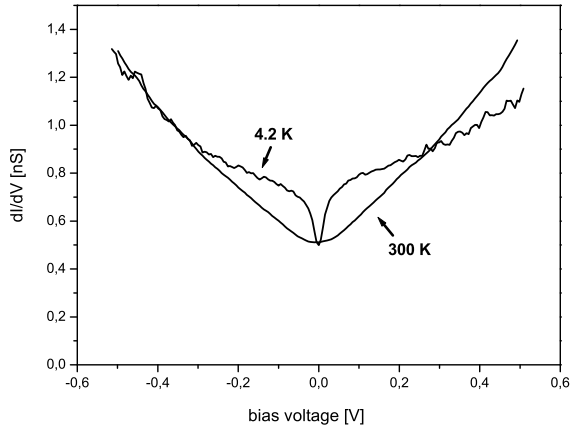}
  				}
\end{center}
\caption{Typical tunnel spectra of FeSi taken with Pt-Rh tip at 4.2~K~and~300~K.}
\label{fig1}
\end{figure}	 	
		  		 		 
 \section{RESULTS AND DISCUSSION}		  		
\begin{figure}
\begin{center}
\resizebox{0.99\columnwidth}{!}{%
  				\includegraphics{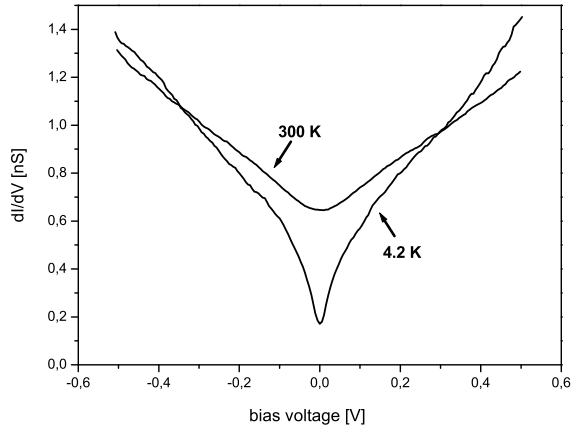}
  				}
\end{center}
\caption{Typical spectra of the FeSi-FeSi tunnel junction at 4.2~K~and~300~K.}
\label{fig2}
\end{figure}

Typical differential conductance curves taken with  
		  \mbox{Pt-Rh} tip 
   		at 4.2~K and 300~K are shown in Fig. \ref{fig1}.
As temperature was lowered from 300~K to 4.2~K, 
			in the region of $\left|V\right| < $ $\sim$300~mV
			the differential conductance increased 
			except the vicinity of the zero-bias, 
			where a	strong and sharp dip was formed, with the minimum 
			of (almost) the same value as the zero bias conductance 
		  at 300~K.
(Here should be noted that there is a relatively strong dependence 
 		 		of the zero bias conductance on the width of the interval used 
 		 		for calculation of numerical derivation.
We used the widest 	possible interval still not increasing 
 		 	 the zero bias conductance; this is the reason for a bigger error
 		 		 of $dI/dV$ curve at 4.2~K in Fig.~\ref{fig1}.)
Exemplary spectra of the FeSi-FeSi configuration 
			are depicted in Fig.~\ref{fig2}.
The conductances observed at 300~K seem to be very similar 
			to those measured out with Pt-Rh tip.
On the other hand, the spectra taken at 4.2~K exhibit 
			qualitatively different features, 
			as they show a visible decrease of the differential conductance 
			under the values of the 300~K curve 
			in the whole energy interval corresponding to 
			\mbox{$\left|V\right| < $ $\sim$300~mV.}
The zero-bias conductance shows the greatest decrease, 
			but at 4.2~K it remains still finite.

According to Hammers \cite{Hammers}, 
				the differential conductance $dI/dV$ in 
				the tip-sample configuration can be expressed by 
\begin{eqnarray}   
{\frac{dI}{dV}=\rho_{s}(r,eV)\rho_t(r, 0)T(eV, eV, r)}
\nonumber \\
+\int_{0}^{eV}\rho_{s}(r,E)\rho_{t}(r,\pm {eV}, \mp E)\frac{dT(E,eV,r)}{dV}dE,
 	\label{eq2}
	 \end{eqnarray}
				where $\rho_s(r, E)$ and $\rho_t(r, E)$ are the state densities of the sample and the tip, 
				respectively, at location $r$ and at the energy $E$, measured with respect
 				to their individual Fermi levels; 
 				$T(E,eV,r)$ is the tunneling transmission probability 
 				for electrons with energy $E$ at applied bias voltage $V$.
The upper and the lower signs in equation (\ref{eq2}) correspond to a positive   
				and a negative sample bias, respectively \cite{Hammers}.
At any fixed location, $T(E,eV,r)$ increases monotonically 
		with $V$ and contributes by a smoothly varying ``background'', 
		on which the spectroscopic information is superimposed \cite{Hammers}.
Because of the smooth and monotonic increase, 
			a structure in $dI/dV$ as a function of~$V$ usually can  
			be assigned to changes in the state density 
			via the first term of equation~(\ref{eq2}), 
			thus permitting the DOS to be determined 
			as a function of energy at any particular 
			location on the surface \cite{Hammers}.
Analogously, if DOS changes due to a variance 
			of temperature, the change in DOS at constant energy $E=eV$ 
			is correspondingly detected as the change of $dI/dV$ 
			at the voltage $V$. 
The change of the DOS due to the temperature variance 
			from $T_1$ to $T_2$  can be then inferred from the difference 
			$dI/dV(T_2) - dI/dV(T_1)$. 
(Of course, in such a case the effects due to temperature 
			smoothing are neglected.
Because of the energy scale, which is a
			 few thousand Kelvins in our case, such an approximation seems to be acceptable.)
		
\begin{figure}
\begin{center}
\resizebox{0.99\columnwidth}{!}{%
  				\includegraphics{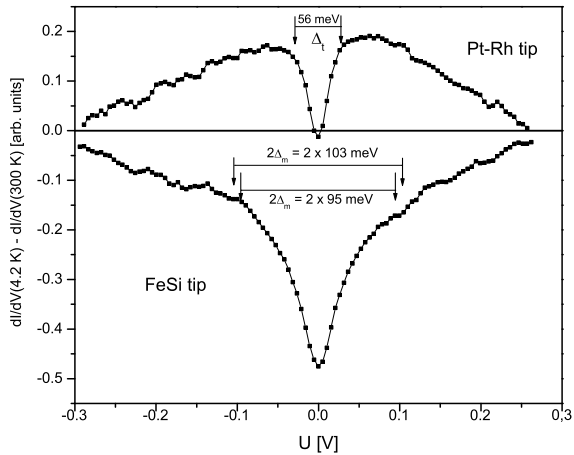}
  				}
\end{center}
\caption{Change of the differential conductance after cooling down
 from 300~K to 4.2~K for Pt-Rh tip (top) and for FeSi-FeSi tunnel junction (bottom).}
\label{fig3}
\end{figure}

Fig.~\ref{fig3} shows the difference between the  $dI/dV$ curves 
			taken at 300~K and 4.2~K
			for both TJ configurations. 
The curve at the top, corresponding to the contacts 
	 	with \mbox{Pt-Rh} tip, shows two almost symmetrical 
	 	local maxima at both sides of the Fermi level with
  	estimated \mbox{peak-to-peak} distance of about 110~meV, 
  	separated by the dip with the minimum 
  	of practically the zero-value at the \mbox{zero-bias}.
Because of the temperature independent DOS of \mbox{Pt-Rh} tip, 
			the observed influence of the temperature on the tunnel spectra 
			should reflect changes in the DOS of FeSi only.
In spite of the fact that the obtained curve can not be
		 without a renormalization quantitatively
			related to the DOS, it clearly indicates 
			a formation of two symmetrically placed peaks at 
 			gap edges. 
(According to the finite values of the differential conductance 
			in  all the studied cases, it would be probably more appropriate to speak about
			\emph {the direct and the indirect ``pseudogap''} in FeSi. Nevertheless, for the purposes of this paper we will use the term 
			 \emph{``gap''}.)		
The difference curve for \mbox{FeSi-FeSi} configuration at the 
 	  		bottom of Fig. \ref{fig3}, is negative 
				in the whole region of $\left|V\right| < $ $\sim$300~mV
				with the strong dip centered at the zero bias. 
Such a behavior can be attributed to a decrease of the quasiparticle
			 DOS 
			 around the Fermi level by falling the 
			 temperature from the room one to 4.2~K.
As follows from the subsequent discussion, we have associated 
				the curve at the top and at the bottom of  Fig. \ref{fig3}
			 with the temperature change of the \emph{d-} and the \emph{c-}partial DOS, respectively. 
			
As emphasized by Tromp \cite{Tromp}, 
			because of the determination of the
			tunneling current by the tunneling transmission probability and by the DOS of both electrodes, 
 			in addition to a finite DOS 
			there must be a significant overlap between the corresponding sample 
			and tip wave functions	 \cite{Tromp}.
If the sample has a large DOS, 
			but these states do not overlap with the tip, 
			they are inaccessible in the tunneling experiment \cite{Tromp}. 

In the case of the heterogeneous TJs with the metallic Pt-Rh tip, 
			a sufficient overlap of the  rather spatially 
			extended wave function 	of conduction electrons of the  tip 
			with that of the correlated \emph{c-} and \emph{d-}electrons in FeSi can be expected. 
Because of much larger and peaked \emph{d-}partial DOS		
			a  major tunnel current will originate from \emph{d-}electrons.
In fact, the shape of the curve at the top of Fig. \ref{fig3} 
				strongly resembles the experimental results of other authors \footnote{see Fig. 3 in \cite{Fath}}
				and the situation in \emph{d-}partial DOS proposed by the
				theory of Rozenberg, Kotliar and Kajueter \cite{Rozenberg}. 
Therefore, in agreement with 
				 	 F\"ath and coworkers \cite{Fath}, 
			we conclude that the tunneling spectroscopy measurements on 
			the tunnel junctions with  the counter electrode from the metal without electron correlation effects
			 probe preferably the much larger and peaked \emph{d-}partial DOS in FeSi.
			
In replacing the \mbox{Pt-Rh} tip  by FeSi tip, 
		the relatively slowly decaying wave function 
		of conduction electrons of \mbox{Pt-Rh} tip 
		is then replaced by the wave functions of correlated  
		\emph{c-} and \mbox{\emph{d-}electrons} of FeSi. 
Due to the localized nature of \mbox{\emph{d}-electrons}
		their wave function is less spatially extended
		than one of \mbox{\emph{c-}electrons}, so it is reasonable 
		to expect that an overlap  between 
		  the \mbox{\emph{c-} states} will be dominating here.
In spite of the high density of \emph{d-} states, the tunneling process in
			\mbox{FeSi-FeSi} type of TJ seems to be predominantly governed by the
			lower-density \emph{c-}electrons, as the effect of the overlap between the 
			 \emph{c-} states seems to be superior 
			to the effect of high density of \emph{d-} states.
In addition to this, the shape of the curve at the bottom of Fig. \ref{fig3}
 			resembles the theoretically predicted change of the 		
			\emph{c-}partial DOS due to temperature variation in the region of the direct gap \cite{Rozenberg}.	
Based on the given arguments we conclude that the
			FeSi-FeSi tunneling configuration yields preferably the
			spectroscopic information on \emph{c-}partial DOS.

According to the previous discussion, 
		the dip observed in TJs with Pt-Rh tip
		is associated with the indirect gap. 
Although the shape of the dip is modified due to Shottky barrier effect \cite {Fath},
		its width should still correlate with 
		the width of the indirect gap. 
As $\Delta_{ind}$ is associated with transport measurements 
	\cite{Paschen,Rozenberg}, 	it is to be expected that the width of this dip
		will be comparable with the transport gap value $\Delta_{t}=56~meV$ previously 
		derived for the studied sample from the resistivity data \cite{Mihalik}. 
In Fig. \ref {fig3}, where  $\Delta_t$ is correspondingly indicated, a definite 
		correlation between $\Delta_{t}$  and  the width of the dip can be seen.
On the other hand, the magnetic gap is related to the direct gap $\Delta_{dir}$
		 \cite{Paschen,Rozenberg}.		 
The width of the magnetic gap of our sample 
		 $\Delta_{m}\cong$ 95~(or~103~meV, depending on the  fitting procedure used)
		 \cite{Mihalik} is larger than $\Delta_{t}$ in accordance with 
		 the theoretical predictions \cite{Rozenberg}.
The comparison of $\Delta_{m}$ with the width of the dip visible in the 
			curve at the bottom of Fig. \ref{fig3} 	 requires taking into account 
 			that in FeSi-FeSi type of TJ there are gap structures at both sides of the TJ.
So that (analogously like it is in superconductor-insulator-superconductor TJs), 
			the decrease of the differential conductance 
			 should develop in the energy interval of $2\Delta_{m}$.			
Doubled values of $\Delta_{m}$ are indicated in Fig. \ref{fig3}.
Although the determination of the dip width is not straightforward, 
			 it can be seen that the indicated values of $2\Delta_{m}$ lie in the crossover 
			 between the dip region and the background region, and so, it can be said that the 				width of the dip region correlates with the $\Delta_{m}$. 
The observed good correspondence between our tunneling data and the earlier results of 					the transport and magnetic studies, provides a further support for the 										interpretation of the tunneling data given above.

\section{CONCLUSIONS}			
Our tunneling spectroscopy studies of FeSi have revealed that two types of electron subsystems 
		and two different energy gaps 
		 are present in FeSi.
Depending on the used type of the counter electrode,
		 the lowering of temperature from 300~K to 4.2~K causes either the formation of the        peaks  in the partial DOS, which we associate with 
		 the properties of \emph{d-}electrons and with the indirect gap; 
		 or decrease of the partial DOS  at  and in the vicinity of the Fermi level, what we
		 assign to the direct gap formation in the \emph{c-}partial DOS.  
The obtained results strongly support the theoretical
		model of Rozenberg, Kotliar and Kajueter \cite{Rozenberg} and  correlate well
		with the independent evaluation of  the transport  
	 and  the magnetic gap of our FeSi sample.
Moreover, our work  shows that 
		the	type of spectroscopic information 
		obtained by tunneling spectroscopy can be ``selected'' by the type of counter 						electrode used. 
For the more detailed information on the role of the counter electrode material and on 				the properties of correlated electron subsystems in FeSi, additional experiments are going to be done at various temperatures, utilizing several types of counter electrode materials.

%

\section*{ACKNOWLEDGEMENTS}
 This work was supported by
the VEGA Grant 2/4050/04,
the Project APVT-51-031704,
and the
Centre of Low Temperature Physics
operated as the Centre of Excellence of the Slovak Academy
of Sciences under Contract No. I/2/2003.
The low-temperature STS-head and STS control
electronics  were provided by Laboratories of Applied Research 
\footnote{Laboratories of Applied Research, Atletick\'a 16, 040 01 Ko\v {s}ice, Slovakia}.

\end{document}